# Comment on "Microscopic magnetic modeling for the S = 1/2 alternating chain compounds $Na_3Cu_2SbO_6$ and $Na_2Cu_2TeO_6$" (arXiv: 1402.1091)

Masatoshi Sato[1]* and Yoshiaki Kobayashi[2]

[1]*Research Center for Neutron Science and Technology, Comprehensive Research Organization for Science and Society, 162-1 Shirakata, Tokai 319-1106 Japan*
[2]*Department of Physics, Nagoya University, Furo-cho, Chikusa-ku, Nagoya 464-8602, Japan*
(Dated: February 27, 2014)

In a recent paper by Schmitt et al. (arXiv: 1402.1091.), signs of two exchange interactions of one-dimensional alternating chains in distorted honeycomb systems of $Na_3Cu_2SbO_6$ and $Na_2Cu_2TeO_6$ are argued by theoretical calculations. Although the authors report that they have clarified that the interaction $J_{1a}$ is ferromagnetic ($J_{1a}<0$) between the spins within a structural dimer, and $J_{1b}$ is antiferromagnetic ($J_{1b}>0$) between these dimers, these results have been known for a long time by our two papers, which are cited by Schmitt et al.

The authors of Ref. [1] (Schmitt et al.) have reported that their calculations evidenced that the magnetism of the spin-gap systems of $Na_3Cu_2SbO_6$ and $Na_2Cu_2TeO_6$ can be described by the alternating Heisenberg chain model with two relevant exchange couplings: ferromagnetic $J_{1a}$ (<0) between the spins within a structural dimer, and antiferromagnetic $J_{1b}$ (>0) between these dimers. In their arguments, they cited our papers [2], [3] (their Refs. [14] and [39], respectively), and compared their calculated results with our data, obtaining the above results. However, we are afraid that it may mislead readers, because of the following reasons.

(1) In our first paper [2], in which we observed the spin-gap phenomenon in these systems, we could not distinguish which of these two exchange interactions ($J_{1a}$ and $J_{1b}$) was ferromagnetic, although the possibility of $J_{1a} \times J_{1b}>0$ was completely ruled out in this work. The authors of Ref. [1] emphasize this point, as if the sign problem still existed for these systems, although we solved the problem in our second paper [3] as explained in (2).

(2) To solve the sign problem of $J_{1a}$ and $J_{1b}$, we studied the $Q$ dependence of the dynamical structure factor $S(Q, \omega)$ by neutron magnetic scattering measurements, and unambiguously determined as $J_{1a}<0$ (ferromagnetic) and $J_{1b}>0$ (antiferromagnetic), where $\omega$ is the excitation energy and $Q$ is the scattering vector expressed as $ha^*+kb^*+lc^*$ with $a$ and $b$ being within a honeycomb plane ($b$||chain direction).

At $\omega=9$ meV close to the minimum spin-gap energy (see Figs. 4 and 5(a) of Ref. [3]), we observed, for $h=1.0$, the intensity peaks of the triplet excitation at $k$ =0.5, 2.5, and 3.5 in the region of $k<4.0$. For $h=1.5$ and 2.0, peaks were found at the same $k$ points as for $h=1.0$. We did not observe any peaks in the $k$-scan profiles along (1, $k$, 0), (1.5, $k$, 0), and (2, k, 0) at $k =1.0$. One dimensional nature of the spin system was also established.

Considering that the static ordering pattern expected, if the spin-gap formation were absent, from their exchange interactions, is reflected in the distribution of $S(Q, \omega)$ in the spin-gap phase, we can safely conclude that the spin correlation should be as shown in Fig. 2(c) of Ref. [3], indicating that $J_1<0$ (ferromagnetic) and $J_{1b}>0$ (antiferromagnetic). We emphasized it in §3 and §4.

The observed dispersion curve of the spin gap energy along the $k$ direction (see Fig. 5(a) in Ref. [3]) explicitly indicates that the minimum spin-gap energy is located at the $Q$-points corresponding to the $S(Q, \omega)$ maximum at $\omega \sim$ the minimum spin-gap energy (~9 meV). The dispersion curve is also reproduced in Ref. [1].

Based on the above points, we would like to stress that the sign problem of $J_{1a}$ and $J_{1b}$ was settled by Refs. [2] and [3]. Of course, the work reported in Ref. [1] has confirmed these experimental results theoretically.